\documentclass[aps,prl,reprint,amsmath,superscriptaddress]{revtex4-2}
\usepackage{graphicx}
\usepackage{amsfonts,amsmath,amssymb}
\usepackage[colorlinks=true]{hyperref}
\usepackage{pgf}
\usepackage{pgfplots}
\usepackage{bm}
\usepackage{comment}

\begin{document}

\title{Geometric Spin Rotation in Triangular Antiferromagnets}

\author{Grigor Adamyan}
\affiliation{William H. Miller III Department of Physics and Astronomy, Johns Hopkins University, Baltimore, Maryland 21218, USA}
\affiliation{Department of Physics, Massachusetts Institute of Technology, Cambridge, Massachusetts 02139, USA}

\author{Basti\'an Pradenas}
\affiliation{William H. Miller III Department of Physics and Astronomy, Johns Hopkins University, Baltimore, Maryland 21218, USA}
\affiliation{Leibniz Institute for Solid State and Materials Research (IFW Dresden), Helmholtzstr. 20, 01069 Dresden, Germany}

\author{Boris Ivanov}
\affiliation{William H. Miller III Department of Physics and Astronomy,
Johns Hopkins University, Baltimore, Maryland 21218, USA}
\affiliation{V.G. Baryakhtar Institute of Magnetism of the National Academy of Sciences of Ukraine, 03142 Kyiv, Ukraine}

\author{Oleg Tchernyshyov}
\affiliation{William H. Miller III Department of Physics and Astronomy,
Johns Hopkins University, Baltimore, Maryland 21218, USA}

\begin{abstract}
We describe a geometric phenomenon in which a traveling wave made of degenerate Goldstone modes leaves behind a transformed ground state. In a triangular Heisenberg antiferromagnet, a pulse of circularly polarized spin waves rotates the spins within their plane. An exact solution of the nonlinear equations of motion demonstrates that the accumulated rotation is a geometric phase related to parallel transport of the order parameter. We point out a curious analogy between the motion of the magnetic order parameter and that of a wobbling coin. This phenomenon opens a new route for controlling antiferromagnetic order by spin waves and may extend to other frustrated magnets as well as other physical systems with noncommuting broken-symmetry generators.
\end{abstract}

\maketitle

Geometric phases arise when a system undergoes cyclic evolution, acquiring a holonomy determined solely by the geometry of the path. An early manifestation is the Aharonov–Bohm effect~\cite{AharonovBohm1959}, while a general formulation in terms of adiabatic transport in parameter space was introduced by Berry~\cite{Berry1984phase}. Since then, geometric phases have emerged as a unifying principle across diverse areas of physics. Classical realizations include the Hannay angle in integrable systems~\cite{Hannay1985angle}, the precession of Foucault's pendulum~\cite{Foucault1851}, and the rotation of light polarization in curved optical fibers~\cite{Pancharatnam1956,Raymond1986photon}. In condensed-matter systems, Berry curvature in crystal-momentum space underlies anomalous transport, orbital magnetization, quantized Hall responses, and the topological classification of electronic bands~\cite{TKNN1982,Zak1989phase,Niu1999,Niu2010,HasanKane2010,QiZhang2011}.

Adiabatic evolution of parameters $\boldsymbol{\lambda}$ around a closed loop $C$ generates a geometric phase
\begin{align}
    \gamma = \oint_C \mathbf{A}(\boldsymbol{\lambda}) \cdot d\boldsymbol{\lambda},
    \label{eq:berry_phase}
\end{align}
where $\mathbf{A}(\boldsymbol{\lambda})$ is the Berry connection~\cite{Berry1984phase}. This framework was subsequently generalized by Wilczek and Zee to degenerate manifolds, giving rise to non-Abelian geometric phases~\cite{WilczekZee1984}. In the conventional setting, $\boldsymbol{\lambda}$ represents externally controlled variables or system degrees of freedom whose evolution is driven by external fields. Canonical examples are the direction of an applied magnetic field for a spin-$1/2$ particle and the crystal momentum of an electron, respectively~\cite{Wilczek1989geometric_phases}.

In this work, we uncover a distinct geometric mechanism that does not rely on externally driven parameter cycles. Instead, we show that intrinsic excitations of the system can themselves generate geometric phases. In systems where the symmetry group of the order parameter is non-Abelian, degenerate Goldstone modes can be associated with noncommuting generators~\cite{Nambu1960,Goldstone1961,Beekman2019ssb}. Superpositions of such modes can therefore induce a higher-order transformation of the ground state along different symmetry-breaking directions. This transformation is geometric in nature and related to the parallel transport of the order parameter.

We have found a particularly transparent realization of this mechanism in the triangular Heisenberg antiferromagnet, which hosts a pair of degenerate Goldstone modes. These are linearly polarized spin waves in which spins oscillate about an axis parallel to the spin plane. Their superpositions with circular or elliptical polarization generate a net rotation of the ground state thanks to the noncommutativity of order-parameter rotations about different axes. We obtain an exact nonlinear solution for such excitations and show that the accumulated rotation is a geometric phase. For circularly polarized spin waves, we have found a precise analogy between the motion of the order parameter and that of a wobbling coin that provides an intuitive picture of this striking phenomenon.

Although we focus on triangular antiferromagnets, the geometric effect should be common to physical systems with degenerate Goldstone modes associated with noncommuting broken-symmetry generators. This suggests broader relevance to frustrated magnets and other condensed-matter systems. Beyond fundamental interest, this phenomenon is particularly timely in view of recent advances in antiferromagnetic spintronics~\cite{Nakatsuji2015,Jungwirth2016,Baltz2018} as it suggests another mechanism to control antiferromagnetic order by spin waves.

To illustrate this effect, we consider the Heisenberg antiferromagnet on a triangular lattice, which exhibits strong geometric frustration and forms three magnetic sublattices. Its classical ground state adopts a coplanar $120^\circ$ spin order, where the net spin vanishes on each triangle. Long-wavelength excitations around the ground state induce a small net spin, with the relative orientations of the three sublattice magnetizations $\mathbf{m}_1$, $\mathbf{m}_2$, and $\mathbf{m}_3$ remaining nearly rigid. Thus, the low-energy dynamics are reduced to rotations of the magnetization triad, analogous to rigid-body rotations~\cite{Andreev1980,DombreRead1989,Chernyshev2009af,Dasgupta2020af,Pradenas2024spin_frame,Tchernyshyov2024lecture_notes}.

The magnetic order parameter is the spin frame $\hat{\mathbf n}$ defined by three orthonormal vectors $\{\mathbf n_x, \mathbf n_y, \mathbf n_z\}$, the first two of which are linear combinations of sublattice magnetizations and the third, $\mathbf n_z = \mathbf n_x \times \mathbf n_y$, is the vector chirality normal to the spins. Within this framework, the low-energy dynamics are described by the Lagrangian~\cite{Pradenas2024spin_frame}
\begin{align}\label{eq:spin-frame-L}
    \mathcal{L} &= \frac{\rho}{4} 
    \dot{\mathbf n}_i \cdot \dot{\mathbf n}_i 
    -\frac{\mu}{2}\partial_{\alpha} \mathbf{n}_{\beta} \cdot \partial_{\alpha} \mathbf{n}_{\beta}.
\end{align}
Greek indices $\alpha,\beta$ run over $x,y$, while the Latin index $i$ runs over $x,y,z$; $\rho$ is the inertia density, and $\mu$ is the exchange stiffness.

Starting with the ground state in which the spin frame is aligned with a global frame $\hat{\mathbf e} = \{\mathbf e_x, \mathbf e_y, \mathbf e_z\}$, we add a small-amplitude excitation parametrized by a rotation angle $\boldsymbol{\phi} = \phi_i \mathbf e_i$:
\begin{align}\label{eq:small-angles}
    \mathbf n_i
    = \mathbf e_i
    + \boldsymbol{\phi}\times\mathbf e_i
    + \frac{1}{2} \boldsymbol{\phi}
    \times(\boldsymbol{\phi}\times\mathbf e_i)
    + \ldots
\end{align}
Expanding the Lagrangian~\eqref{eq:spin-frame-L} 
to second order in $\boldsymbol \phi$ yields
\begin{equation}\label{eq:L2}
\mathcal{L}_2 
= \frac{\rho}{2} 
\dot{\phi}_\alpha \dot{\phi}_\alpha  
- \frac{\mu}{2} 
\nabla \phi_\alpha \cdot \nabla \phi_\alpha
+ \frac{\rho}{2} \dot{\phi}_z^2 
- \mu (\nabla \phi_z)^2,
\end{equation}
where $\nabla = (\partial_x,\partial_y)$ is the two-dimensional spatial gradient and $\alpha=x,y$ is summed~\cite{Pradenas2024spin_frame,Pradenas2024helical}.

The quadratic theory gives three Goldstone modes with linear dispersion. The degenerate modes $\phi_x$ and $\phi_y$ correspond to out-of-plane oscillations of the magnetization triad, propagate at speed $c = \sqrt{\mu/\rho}$, and can be combined into circularly polarized waves,
\begin{align}\label{eq:circ_pol_SW}
    \phi_x \pm i\phi_y = \frac{\phi_0}{\sqrt{2}} \, e^{i (\omega t - \mathbf{k} \cdot \mathbf{r})}.
\end{align}  
Their quanta carry isospin $I_3 = \pm \hbar$. The third mode, $\phi_z = \phi_0 \cos{(\omega t - \mathbf{k}\cdot\mathbf{r})}$, describes in-plane oscillations, propagates at the higher speed $\sqrt{2} \, c$, and carries isospin $I_3=0$~\cite{Pradenas2025ssb}. 

Spin waves with a circular (or more generally elliptical) polarization can be visualized as superpositions of two linearly polarized spin waves with a phase shift or as rotations about an instantaneous axis that itself keeps rotating in the spin plane. Because the resulting motion involves noncommuting rotations of the order parameter, such superpositions induce a net in-plane rotation of the spins $\phi_z$. Since this effect lies beyond the scope of the quadratic theory~\eqref{eq:L2}, we expand the Lagrangian to the third order in $\boldsymbol \phi$ and find a term describing an interaction between the in-plane and out-of-plane modes,
\begin{align}\label{eq:L3_int}
\mathcal{L}_\text{int} = 
 \frac{\mu}{2} \, \nabla \phi_z \cdot 
 (\phi_x \nabla \phi_y - \phi_y \nabla \phi_x).
\end{align}
The Lagrangian yields the equation of motion for the in-plane $\phi_z$ mode,
\begin{equation}\label{eq:eom_phi_z}
    \rho\, \ddot{\phi}_z - 2\mu\, \nabla^2 \phi_z = -\frac{\mu}{2} (\phi_x \nabla^2 \phi_y - \phi_y \nabla^2 \phi_x),
\end{equation}
where the right-hand side acts as a source generated by the out-of-plane $\phi_x \pm i\phi_y$ modes. 

We seek solutions in the form of a traveling wave $\phi_i(\tau)$, where $\tau = t \pm x/c$. Substituting this Ansatz into Eq.~\eqref{eq:eom_phi_z} yields 
\begin{equation}
\phi_z''
= \frac{1}{2}
(\phi_x \phi_y'' - \phi_y \phi_x''),
\end{equation}
where primes denote the derivatives with respect to $\tau$. This equation can be integrated once to obtain
\begin{equation}
\label{eq:rot_phi_z_rate}
   \phi_z' = \frac{1}{2} 
   (\phi_x \phi_y' - \phi_y \phi_x') 
   + \text{const}.
\end{equation}
For configurations that are static in the deep past and future, the integration constant vanishes, so we obtain
\begin{equation}
d\phi_z 
= \frac{1}{2}(\phi_x\, d\phi_y - \phi_y\, d\phi_x).
\label{eq:d-phi-z}
\end{equation}

The absence of the time argument $\tau$ in Eq.~\eqref{eq:d-phi-z} suggests that the accumulation of the in-plane angle $\phi_z$ is a \emph{geometric effect}. To gain further insight, consider a localized wave packet that begins at $\phi_x = \phi_y = 0$ at $\tau=0$ and returns to the same values at $\tau = T$, tracing a closed loop $\partial \Omega$ that bounds a region $\Omega$ in the $\{\phi_x, \phi_y\}$ parameter space. The increment 
\begin{equation}\label{eq:gamma_Berry}
\Delta \phi_z
    = \oint_{\partial \Omega} \frac{1}{2}(\phi_x\, d\phi_y - \phi_y\, d\phi_x)
    = \int_\Omega d \phi_x \, d \phi_y
\end{equation}
is the area inside the loop by Stokes' theorem. 

In the triangular antiferromagnet, this geometric phase appears as a rotation of the spin frame about the $\mathbf{n}_z$ axis driven by spin waves with both $\mathbf{n}_x$ and $\mathbf{n}_y$ polarizations. When a packet of such spin waves passes through an antiferromagnet, the magnetic order is rotated in its wake.

\begin{figure*}[t]
  \centering
  \includegraphics[width=1\linewidth]{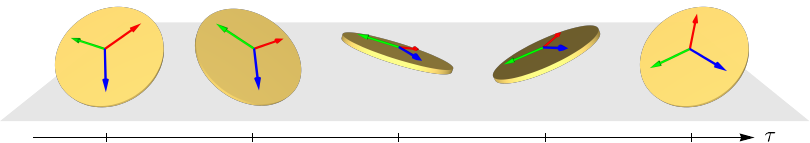}
  \caption{A wobbling coin: a full wobble results in a net rotation of the coin's surface. Analogously, under circularly polarized spin waves, the magnetization triad $\{\mathbf{m}_1, \mathbf{m}_2, \mathbf{m}_3\}$ (red, green, and blue arrows) exhibits the same motion, leaving the antiferromagnet's ground state rotated.}
  \label{fig:wobbling_coin_triad}
\end{figure*}

Notably, we have found an exact analytical solution that exhibits the effect of ground-state rotation in the fully nonlinear regime. To describe it, we parametrize the orientation of the spin frame $\hat{\mathbf n}$ by the Euler angles. Starting from a reference configuration in which the spin frame $\hat{\mathbf n}$ is aligned with a fixed global frame $\hat{\mathbf e}$, we perform three consecutive rotations: by $\Phi$ about $\mathbf{n}_z$, then by $\Theta$ about $\mathbf{n}_x$, and finally by $\Psi$ about $\mathbf{n}_z$. The Euler angles $\Phi$, $\Theta$, and $\Psi$ serve as dynamic fields encoding the local orientation of the spin frame. The Lagrangian of the triangular antiferromagnet~\eqref{eq:spin-frame-L} takes on the form
\begin{eqnarray}
\mathcal{L} &=& 
\frac{\rho}{2}
\big[\dot{\Theta}^2 
+ \dot{\Phi}^2 \sin^2\Theta
+ (\dot{\Psi} + \dot{\Phi} \cos\Theta )^2\big] 
\label{eq:L_Euler}
\\
&-& \frac{\mu}{2}
\big[(\nabla \Theta)^2 
+ (\nabla \Phi)^2 \sin^2\Theta 
+ 2(\nabla \Psi + \nabla\Phi \cos\Theta)^2\big].
\nonumber
\end{eqnarray}

As in the perturbative solution, we seek a traveling-wave solution $\Phi(\tau)$, $\Theta(\tau)$, $\Psi(\tau)$ with $\tau = t - x/v$. The velocity $v$ of the traveling wave is to be determined. Substituting this Ansatz into Eq.~\eqref{eq:L_Euler} yields the Lagrangian of a symmetric top~\cite{LL1},
\begin{align}\label{eq:L_Euler_prop_wave}
    \mathcal{L} = \frac{I_1}{2} \big(\dot{\Theta}^2 + \dot{\Phi}^2 \sin^2{\Theta} \big) +
    \frac{I_3}{2} \big(\dot{\Psi} + \dot{\Phi} \cos{\Theta}  \big)^2,
\end{align}
with moments of inertia 
\begin{equation}
I_1 = \mu(c^{-2} - v^{-2}), 
\quad
I_3 = \mu(c^{-2} - 2v^{-2}).
\end{equation}

Translational symmetries in $\Phi$ and $\Psi$ yield  conserved momenta
\begin{align}
    \begin{aligned}
        p_\Psi &= \frac{\partial \mathcal{L}}{\partial \dot{\Psi}} 
               = I_3 \big(\dot{\Psi} + \dot{\Phi} \cos{\Theta} \big), \\
        p_\Phi &= \frac{\partial \mathcal{L}}{\partial \dot{\Phi}} 
               = I_1 \, \dot{\Phi} \sin^2{\Theta} + p_\Psi \cos{\Theta}.
    \end{aligned}
\end{align}
These are projections of the angular momentum onto the local axis $\mathbf{n}_z$ and global axis $\mathbf{e}_z$, respectively. 

The values of the conserved momenta are determined by boundary conditions. For a localized traveling wave, the system approaches its ground state far from the excitation core, so $\Theta \rightarrow 0$ as $\tau \rightarrow \pm \infty$. In these limits, the orientation of the spin frame is described by a single rotation angle $\Phi + \Psi$ about the axis $\mathbf{n}_z = \mathbf{e}_z$. Therefore, we also require $\dot{\Phi} + \dot{\Psi} \to 0$ as $\tau \to \pm \infty$. We thus obtain
\begin{equation}
p_\Psi = 0, 
\quad 
p_\Phi = 0.  
\label{eq:p-Psi=p-Phi=0}
\end{equation}

For $p_\Phi = 0$, a nontrivial solution with $\Theta \neq 0$ requires the vanishing of the moment of inertia $I_1$. Although that would be unphysical for a spinning top~\cite{LL1,Goldstein1980mechanics}, in our problem setting $I_1 = 0$ merely fixes the propagation speed at $|v| = c$. 

The vanishing of conserved momentum $p_\Psi$ \eqref{eq:p-Psi=p-Phi=0} allows us to relate the increments of $\Psi$ and $\Phi$: 
\begin{align}\label{eq:p_psi=0}
    d\Psi = - \cos\Theta\, d \Phi,
\end{align}
with the time variable $\tau$ dropping out once again. 

In a localized traveling wave, where $\Theta \to 0$ outside of the pulse, the orientation of the spin frame in the deep past and future is determined by the sum $\Phi + \Psi$: it is the angle of rotation about the axis $\mathbf n_z = \mathbf e_z$. We thus find the angle of rotation in the spin plane after the propagation of the traveling wave: 
\begin{align}\label{eq:gamma_Euler}
    \gamma = (\Phi + \Psi) \Big|_{-\infty}^{+\infty} = \oint_{\partial \Omega} ( 1 - \cos\Theta ) \, d\Phi.
\end{align}
The rotation angle is equal to the solid angle subtended by the trajectory $\partial \Omega$ of the vector $\mathbf{n}_z$. Equivalently, $\gamma$ is the signed area enclosed on the unit sphere by the trajectory $\partial \Omega$. In the small-amplitude limit, $\Theta \ll 1$, the trajectory remains confined to a small region where the unit sphere is approximately flat. In this regime, the area on the sphere reduces to the planar area in the $\{\phi_x, \phi_y\}$ space, recovering the approximate result~\eqref{eq:gamma_Berry}.

The geometric rotation can be cast in the standard Berry-phase form,~\eqref{eq:berry_phase}, by taking the parameter to be the vector chirality $\boldsymbol{\lambda}=\mathbf{n}_z$. In this representation,
\begin{align}
\gamma = \oint_{\partial \Omega} \mathbf{A}(\mathbf{n}_z) \cdot d\mathbf{n}_z,
\qquad
\mathbf{A}(\mathbf{n}_z)
= \frac{\mathbf{n}_z \times \mathbf{m}}{1 - \mathbf{n}_z \cdot \mathbf{m}},
\end{align}
where $\mathbf{A}(\mathbf{n}_z)$ is the Berry connection of a unit-charge Dirac monopole~\cite{Dirac1931,WuYang1975,Nakahara2018geometry}. In the parametrization~\eqref{eq:gamma_Euler}, the associated Dirac string is aligned with 
$\mathbf{m}=-\mathbf{e}_{z}$. A change of reference frame $\hat{\mathbf e}$ corresponds to gauge transformations, which move the string to an arbitrary direction, while leaving the physical phase $\gamma$ unchanged.

Remarkably, the nonlinear motion of the magnetization triad driven by circularly polarized spin waves mirrors the kinematics of a \emph{wobbling coin}~\cite{Routh1905treatise,Moffatt2000euler_disk}, as illustrated in Fig.~\ref{fig:wobbling_coin_triad}. Drawing an analogy with the antiferromagnet, we define the instantaneous body-fixed orthonormal frame of the coin as $\hat{\mathbf n}=\{\mathbf n_x, \mathbf n_y, \mathbf n_z\}$, where $\mathbf{n}_z$ is normal to the coin's surface. The no-slip constraint requires that the instantaneous angular velocity lie along the line joining the contact point to the center of mass. As a result, the angular velocity $\boldsymbol{\omega}$ has no component along the $\mathbf{n}_z$ axis:
\begin{align}\label{eq:omega_3}
    \boldsymbol{\omega} \cdot \mathbf{n}_{z}  = \dot{\Psi} + \dot{\Phi} \cos{\Theta} = 0,
\end{align}
which reproduces Eq.~\eqref{eq:p_psi=0}. This constraint determines the total rotation angle $\gamma$~\eqref{eq:gamma_Euler} accumulated by the coin's surface during the wobbling motion.

Both the spin frame of the antiferromagnet and the wobbling coin accumulate a net rotation about $\mathbf{n}_z$ over time, despite having no instantaneous angular velocity along this axis. This behavior stems from the \emph{noncommutativity} of $\mathrm{SO(3)}$ rotations. To illustrate, consider a sequence of four infinitesimal rotations: (1) about axis $\mathbf{n}_x$ by infinitesimal angle $\alpha$, (2) about $\mathbf{n}_y$ by $\beta$, (3) about $\mathbf{n}_x$ by $-\alpha$, (4) about $\mathbf{n}_y$ by $-\beta$,
\begin{align}\label{eq:R4}
    \hat{R} = \hat{R}_y(-\beta) \hat{R}_x(-\alpha) \hat{R}_y(\beta) \hat{R}_x(\alpha),
\end{align}
where $\hat{R}_i(\alpha) = 1-i \alpha \hat{L}_i + \ldots$ and $\hat{L}_i = \hat{\bm{L}} \cdot \mathbf{n}_i$ is the generator of rotations about the local $\mathbf{n}_i$ axis. Expanding to the lowest non-vanishing order in $\alpha$ and $\beta$ yields
\begin{align}
    \hat{R} = 1 + \alpha \beta \, [\hat{L}_x, \hat{L}_y] + \ldots
    = \hat{R}_z(\alpha \beta),
\end{align}  
where we have used the commutation relation
\begin{equation}
[\hat{L}_x,\hat{L}_y] = -i \hat{L}_z.   
\end{equation}
It differs in sign from the standard relations for generators of rotations about the global $\mathbf{e}_i$ axes~\cite{LL3}.

\begin{figure}[t]
  \centering
  \includegraphics[width=0.45\linewidth]{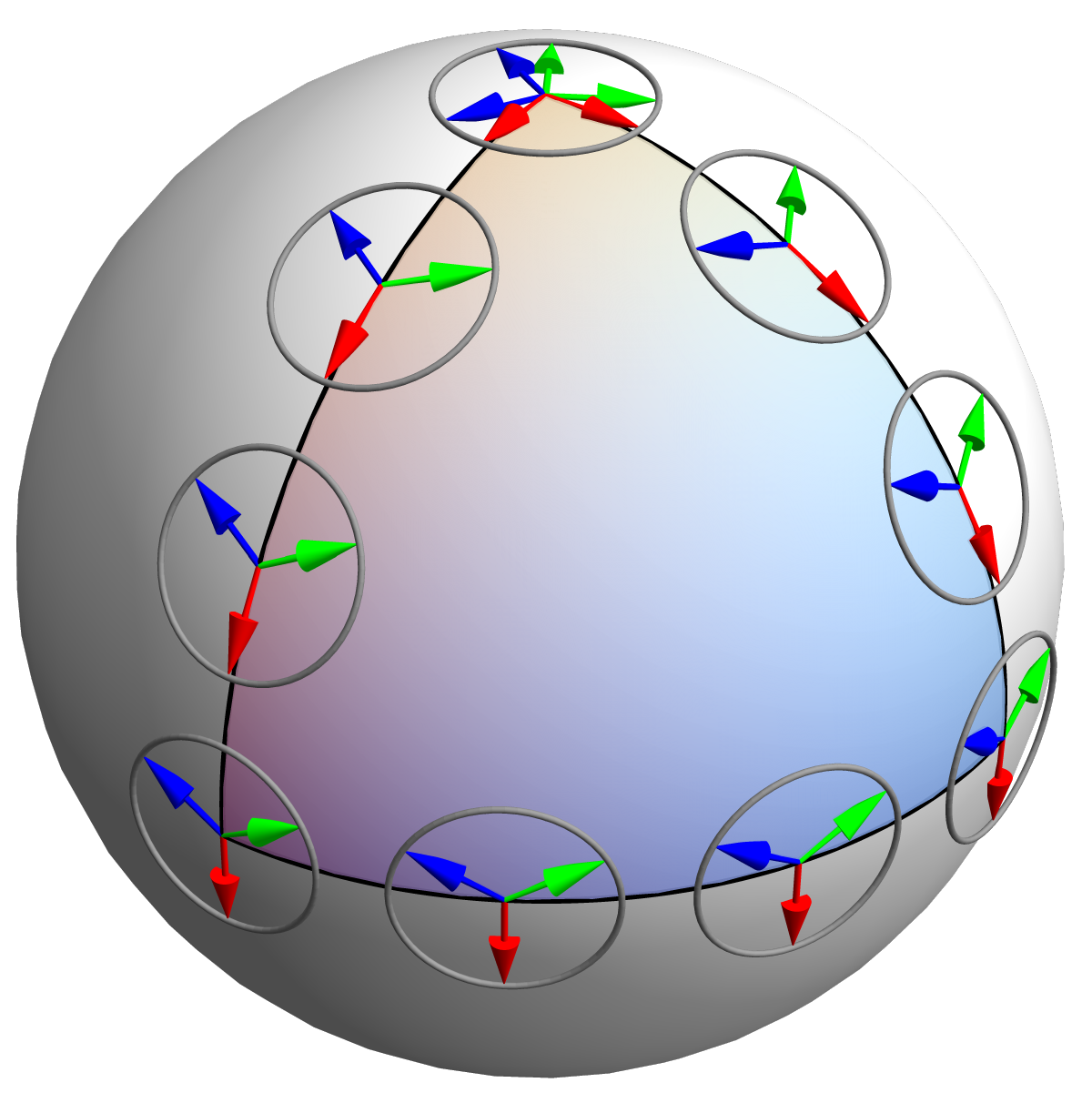}
  \caption{Parallel transport of the magnetization triad on the unit sphere. After completing a loop, the triad is rotated relative to its initial orientation by the angle $\gamma$ equal to the solid angle $\Omega$ enclosed by the loop. Here $\gamma = \Omega = \pi/2$.}
  \label{fig:parallel_transport_triad}
\end{figure}

Geometrically, the spin frame vectors $\mathbf{n}_x$ and $\mathbf{n}_y$ (equivalently, the magnetization triad) undergo \emph{parallel transport} on the unit sphere~\cite{Berry1990,Wilczek1989geometric_phases}. The absence of instantaneous angular velocity about the $\mathbf{n}_z$ axis~\eqref{eq:omega_3} is precisely the parallel-transport condition. Consequently, when the triad traces a closed loop on the sphere, it returns rotated by a net angle $\gamma$~\eqref{eq:gamma_Euler}, equal to the solid angle enclosed by the path (Fig.~\ref{fig:parallel_transport_triad}). This effect underlies such familiar phenomena as the precession of Foucault's pendulum and the rotation of light polarization in curved optical fibers~\cite{Foucault1851,Pancharatnam1956,Raymond1986photon}. The motion of the magnetic order parameter in triangular antiferromagnets offers another example of parallel transport.

In summary, we have suggested a novel geometric mechanism in which superpositions of degenerate Goldstone modes induce a transformation of the ground state. In triangular antiferromagnets, this effect manifests as a net rotation of the spin frame $\hat{\mathbf n}$ driven by polarized spin waves with circular or elliptical polarization. An exact nonlinear solution confirms that even though such waves produce no instantaneous rotation about the $\mathbf{n}_z$ axis, the noncommutativity of rotations about other axes orthogonal to it results in a net $\mathbf{n}_z$ rotation.

In practice, excitation of nonlinear spin waves and monitoring of the ground state can be achieved through the optical pump-probe method~\cite{Kirilyuk2010}. This approach can produce spin waves with varying polarizations traveling at the characteristic spin-wave velocity~\cite{Hortensius2021} in various antiferromagnets, including three-sublattice ones~\cite{Tzschaschel2019}.

The geometric effect discussed here may extend beyond triangular antiferromagnets to other frustrated lattices, including kagome, pyrochlore, and face-centered cubic systems. More broadly, the phenomenon is expected to occur in other systems with noncommuting broken-symmetry generators. Notable examples include spin-triplet superfluids and biaxial nematic liquid crystals~\cite{Volovik2009,ChaikinLubensky1995}.

We thank Hua Chen and Satoru Nakatsuji for stimulating discussions. Research was supported in part by the U.S. Department of Energy, Basic Energy Science, Award No. DE-SC0009390, and the Army Research Office under Award Number W911NF-24-1-0152. G. A. acknowledges support of the Miller Graduate Fellowship at JHU and the Physics Department Fellowship at MIT.

\bibliography{references}

\end{document}